\documentstyle[12pt]{article}
\def\beq{\begin{equation}}
\def\eeq{\end{equation}}
\def\bea{\begin{eqnarray}}
\def\eea{\end{eqnarray}}
\def\bq{\begin{quote}}
\def\eq{\end{quote}}

\def\nnb{\nonumber}

\def\rar{\rightarrow}
\def\nnb{\nonumber}
\def\la{\langle}
\def\ra{\rangle}
\def\nin{\noindent}
\begin{document}
\topmargin -1.5cm
\oddsidemargin +0.2cm
\evensidemargin -1.0cm
\pagestyle{empty}
\begin{flushright}
{CERN-TH.7188/94}\\
PM 94/08
\end{flushright}
\vspace*{5mm}
\begin{center}
\section*{\bf
SIMPLE BUT EFFICIENT TESTS \\
OF QCD FROM TAU DECAYS$^{*)}$}
\vspace*{0.5cm}
{\bf S. Narison} \\
\vspace{0.3cm}
Theoretical Physics Division, CERN\\
CH - 1211 Geneva 23\\
and\\
Laboratoire de Physique Math\'ematique\\
Universit\'e de Montpellier II\\
Place Eug\`ene Bataillon\\
34095 - Montpellier Cedex 05\\
\vspace*{1.5cm}
{\bf Abstract} \\ \end{center}
\vspace*{2mm}
\noindent
We review the determinations of the QCD coupling $\alpha_s$
from the inclusive and exclusive
modes of $\tau$-decays.
The most recent $\tau$-data provide the average value  $
\alpha_ s \left(M^2_\tau \right)\simeq 0.347\pm 0.030 $
corresponding to $ \alpha_ s \left(M^2_Z \right) \simeq  0.121\pm 0.003
\pm 0.001.$
The values of the QCD
vacuum condensates extracted from the inclusive weighted-moment
distributions
are consistent with the ones from QCD
spectral sum rules. Accurate estimates and further precision
tests should be reached in
the next
$\tau$C/B-factory machines.
The $M_\tau$-stability test
 from the $ e^+e^- \longrightarrow
I=1 $ hadron data is also discussed.

\vspace*{2.0cm}
\noindent
\rule[.1in]{16.0cm}{.002in}

\noindent
$^{*)}$ Talk given at the Symposium {\it ``Symmetry and simplicity in
physics"} on the occasion of {\it Sergio Fubini's 65th birthday}, Turin,
24--26 February 1994 and at the QCD--LEP meeting on $\alpha_s$, Aachen,
11 April 1994.
 To appear in the ``Atti dell'Accademia
delle Science di Torino" and in the Aachen proceedings.
\vspace*{0.5cm}
\noindent


\begin{flushleft}
CERN-TH.7188/94 \\
PM 94/08\\
April 94
\end{flushleft}
\vfill\eject
\pagestyle{empty}

\setcounter{page}{1}
\pagestyle{plain}





\section{INTRODUCTION} \par
\medskip
The $ \tau $ semileptonic decay modes have been shown
\cite{BRA}--\cite{BRAT} to be a
``good
laboratory" for testing the perturbative and non-perturbative aspects
of QCD. In particular, due to their inclusive nature, which is even
greater than in the case of $e^+e^-$ into hadrons,
these decay modes can provide a measurement of the QCD
coupling $ \alpha_ s \left(M_\tau \right) $
with ``unprecedented" accuracy.

\nin
Before going to the content of the talk, let me quote the following
statements,
  which will give the flavour of the subject that I shall discuss:

\nin
{\it  Indeed, it is for this process that more work has been done
concerning higher power corrections than for any other processes. The
non-perturbative condensates apparently give a small contribution thus
making this relatively low-energy process a prime place to determine
the $\Lambda$-parameter of QCD}... (A. Mueller \cite{AMU}).

\nin
{\it Another entry in the table of Figure 3 (see Figs.3 and 4)
 is particularly
interesting and provocative, and suggests some later developments, so I
want to go into a little more detail regarding it. It is the
determination of $\alpha_s$ from QCD with corrections to the tau lepton
lifetime. Tau lepton decay of course is a very low energy process by the
standards of LEP or most other QCD tests. So we can expect, in line with
the previous discussion, that the prediction will perhaps be delicate but
on the other hand it will have a favorable lever arm for determining
$\alpha_s$...}
(F. Wilczek \cite{WIL}).

\nin
{\it Tau-decay is a lucky process...} (G. Veneziano).
\par
\smallskip
\nin
In this talk ({\it desol\'e du peu!}),
I shall review the determinations of $ \alpha_ s $ from the
inclusive and from the sum of the exclusive $ \tau $-decay modes.
I shall also discuss the weighted-moment
distributions for  simultaneously measuring $ \alpha_ s $ and the
non-perturbative QCD
condensates.
The $ e^+e^-
\longrightarrow  I = 1 $
hadron data will be used for testing the stability of the result for
arbitrary values of the
$ \tau $-mass.
\section{ \boldmath{${\bf \alpha}_ {\bf s} \left( {\bf M}_ {\bf \tau}
\right) $}{\bf\ FROM THE INCLUSIVE MODE }}\par
\nin
This section is mainly based on the paper in \cite{BRA2}.
\par
\subsection{The na\"\i ve quark-parton model}
\par
\smallskip
{}From the well-known na\"\i ve quark-parton model, one predicts :
\medskip
\beq
 R_\tau   \equiv  {\Gamma \left(\tau  \longrightarrow  \nu
_\tau + hadrons \right) \over \Gamma \left(\tau  \longrightarrow  \nu_ \tau
e\bar  \nu_ e \right)} = N_c,
\eeq
\par
\smallskip
\noindent
very analogously to
\medskip
\beq
{ R_{e^+e^-}   \equiv  {\sigma \left(e^+e^- \longrightarrow
hadrons \right) \over \sigma \left(e^+e^- \longrightarrow  \mu^ +\mu^ -
\right)} =  N_c \sum^{ }_{  i\equiv u,d,s}Q^2_i ,       }
\eeq
\par
\smallskip
\noindent as the two processes can be related to each other through an $
SU(2)_I
$ rotation.
The present data average from the $ \tau $-lifetime :
\medskip
\beq
{ R^\Gamma_ \tau    \equiv  {\Gamma_ \tau - \displaystyle \sum^{
}_{ e,\mu} \Gamma_{ \tau  \longrightarrow  \ell} \over \Gamma_{ \tau
\longrightarrow  \ell}   } = 3.55 \pm  0.06        }
\eeq
\par
\smallskip
\noindent and from the $ \tau $-leptonic branching ratios :
\medskip
\beq
{ R^B_\tau    \equiv  {1 - B_e- B_\mu \over B_e} = 3.64 \pm  0.03
 }
\eeq
\par
\smallskip
\noindent gives :
\medskip
\beq
  R_\tau  = 3.62 \pm  0.03\ \ \ .
\eeq
\par
\smallskip
\nin
This experimental value is indeed a good evidence for the existence of
colour but it
is 20\% higher than the quark-parton model estimate.
\par
\medskip
\subsection{QCD formulation of the tau decay}
\par
\smallskip
Here, we
propose to
study the different QCD corrections on $ R_\tau $ and show the
ability of QCD to resolve this 20\%
discrepancy. In so doing, we shall be concerned with the decay rate :
\medskip
\beq
{R_\tau = 12\pi \ \int^{ M^2_\tau}_ 0\ {ds \over M^2_\tau} \
 \left(1 - {s
\over M^2_\tau} \right)^2 \left\{ \left(1+{2s \over M^2_\tau} \right)\
{\rm Im}\
\Pi^{ (1)}_{(s)} +
{\rm Im}\ \Pi^{ (0)}_{(s)} \right\} , }
 \eeq
\par
\smallskip
\noindent where Im $ \Pi^{ (J)} $ is the hadronic spectral function of a
hadron of spin $ J. $ In
QCD, these two-point correlators can be expressed as :
\medskip
\beq { \Pi^{ (J)}  = \sum^{ }_{ q\equiv d,s} \left\vert V_{uq}
\right\vert^ 2 \left(\Pi^{ (J)}_{uq,V}+ \Pi^{ (J)}_{uq,A} \right);  }
\eeq
\par
\smallskip
\noindent $ V_{uq} $ is the CKM mixing matrix, and the correlators
\medskip
\bea
  \Pi^{ \mu \nu}_{ ij,V}\equiv  i\ \int^{ }_{ } d^4x\ e^{iqx} \left\langle 0
\left\vert T\ V^\mu_{ ij}(x) \left(V^\nu_{ ij}(0) \right)^+ \right\vert 0
\right\rangle \nnb \\
 \Pi^{ \mu \nu}_{ ij,A}\equiv  i\ \int^{ }_{ } d^4x\ e^{iqx} \left\langle 0
\left\vert T\ A^\mu_{ ij}(x) \left(A^\nu_{ ij}(0) \right)^+ \right\vert 0
\right\rangle \nnb \\
{ \Pi^{ \mu \nu}_{ ij\ V/A}  = - \left(g^{\mu \nu} q^2- q^\mu
q^\nu \right)\ \Pi^{ (1)}_{ij,V/A}+ q^\mu q^\nu \Pi^{ (0)}_{ij\ V/A},
}
\eea
\par
\smallskip
\noindent are associated to the quark vector and axial-vector local currents
\medskip
\beq { V^\mu_{ ij}  \equiv :\bar  \psi_ i\ \gamma_ \mu \ \psi_ j:\ \ \
\ {\rm and} \ \ \
 A^\mu_{ ij}\equiv :\bar \psi_ i\ \gamma_ \mu \ \gamma_ 5\ \psi_ j:\ .
}
\eeq
\par
\smallskip
\noindent It is clear that $ R_\tau $ in (6)
 cannot be calculated directly
from QCD for
$ s\leq \Lambda^ 2 $. However,
exploiting the analyticity of the correlators
$ \Pi^{ (J)} $ and the
Cauchy theorem, one can express $ R_\tau $ as a contour integral in the
complex  $s$-plane running
counter-clockwise around the circle of radius $ \vert s\vert
= M^2_\tau $ (see Fig. 1):
\medskip
\beq{ R_\tau =  6i\pi \ \oint^{ }_{\vert s\vert =M^2_\tau} \ {ds
\over M^2_\tau} \ \left(1-{s \over M^2_\tau} \right)^2 \left\{ \left(1+{2s
\over M^2_\tau} \right)\ \Pi^{ (1)}_{(s)}+ \Pi^{ (0)}_{(s)} \right\} .
} \eeq
\noindent One should notice the existence of the double zero at $ s =
M^2_\tau $, which
suppresses the uncertainties near the time-like axis. As $ \vert s\vert  =
M^2_\tau \gg  \Lambda^ 2 $, one
can use the standard operator product expansion (OPE) \`a la SVZ
\cite{SVZ}
 for
the
estimate of the correlators :
\medskip
\beq{ \Pi^{ (J)}  = \sum^{ }_{ D=0,2,4,...}{1 \over (-s)^{D/2}}\
\sum^{ }_{ dim\ O=D}C^{(J)}(s,\mu )\ \langle O(\mu )\rangle ,} \eeq
\par
\smallskip
\noindent where $ \mu $ is an arbitrary scale that separates the long- and
short-distance
dynamics; $ C^{(J)} $ are the Wilson coefficients calculable in perturbative
QCD, while $\la O \ra$ are the non-perturbative operators.
\subsection{The (non-)perturbative operators $ \langle O\rangle $
}
\par
\smallskip
In this part, we shall limit ourselves to the discussions of
the local gauge-invariant operators
$ \langle O\rangle $
of dimension $ D $ which appear in the standard SVZ-expansion.
Discussions of some
eventual effects not included in this expansion will be done later on.
\par
\smallskip
\noindent $ {\bf - \ \la O_2\ra\equiv
 \bar  m^2_i,\ \bar  m_i\bar  m_j} $ and $\bar m^2_j$ {\bf are
 products of the
running quark masses } of the QCD
Lagrangian which are the only possible dimension-2 local gauge-invariant
operators that can be built from the quark and/or gluon fields.
The values of the quark masses determined from the sum rules are
\cite{SN1}, \cite{SNB}:

\medskip
\beq{\hat  m_u  = (8.7 \pm  1.5)
{\mbox MeV}\ \ \ \hat  m_d= (15.4 \pm
1.5){\mbox MeV}\ \ \ \hat  m_s= (270 \pm  30){\mbox MeV} ,} \eeq
\par
\smallskip
\noindent where the invariant mass $ \hat  m $ is related to the running
quark-mass as
\medskip
\beq{\hat  m_i  = \left(\log
\ M_\tau /\Lambda \right)^{\gamma_1/-\beta 1}\
\bar  m_i \left(M^2_\tau \right)\ \left\{ 1 + O \left({\alpha_ s \over \pi}
\right) \right\} \ \ ,  } \eeq
where $\gamma_1 =2$ and for 3 flavours $ -\beta_1=9/2$ are respectively
the first coefficients of the quark anomalous dimension and of the $\beta
$-function.
\par
\smallskip
\noindent $ {\bf - \la \ O_4\ra
\equiv  m_i \left\langle\bar  \psi_ i\psi_ j
 \right\rangle ,\
\left\langle \alpha_ sG^2 \right\rangle} $ {\bf are the quark and gluon
 condensates}.
The former is
known from PCAC and from the sum rules analysis of the $ SU(3)_F $
breaking.
The second is determined from a combined analysis of the charmonium and $
e^+e^- $
into $ I = 1 $ hadron data. Their values are \cite{SN1}, \cite{SNB}:
\beq {\hat  \mu_ u  = \hat  \mu_ d= (189 \pm  7)
{\mbox MeV}\ \ , \ \ \hat  \mu_ s=
 (160 \pm
10){\mbox MeV}  } \eeq
and \cite{SNB}:
\medskip
\beq{ \left\langle \alpha_ sG^2 \right\rangle    = (0.06 \pm
0.03){\mbox GeV}^4 , } \eeq
\par
\smallskip
\noindent where $ \hat  \mu_ i $ are the invariant condensates defined as :
\medskip
\beq{ \left\langle\bar  \psi_ i\psi_ i \right\rangle
\left(M_\tau \right)   = -\hat  \mu^ 3_i \left(\log\ M_\tau /\Lambda
\right)^{2/-\beta 1} \left\{ 1 + O \left(\alpha_ s \right) \right\} \ \ .
} \eeq
However, due to the operator mixings
 in the massive quarks case, the building of the RGI quark
and gluon condensates
needs the inclusion of a tiny $m^4$ perturbative
terms due to the light quark masses as given in the Appendix B of
\cite{BRA2}.
\par
\smallskip
\noindent $ {\bf - \ \left\langle O_6 \right\rangle   \equiv  \alpha_ s
\left\langle\bar  \psi_ i\Gamma_ 1\psi_ i\bar  \psi_ j\Gamma_ 2\psi_ j
\right\rangle } $ {\bf are
 the dimension-6 operators } in the chiral limit
$ m_i= 0 $; $ \Gamma_{ 1,2} $ are generic notations for any Dirac and colour
matrices. One
can express it as :
\medskip
\beq{\left\langle O_6 \right\rangle  = {1 \over 16 N^2_c}\ \left\{
 T_r\Gamma_
1T_r\Gamma_ 2- T_r \left(\Gamma_ 1\Gamma_ 2 \right) \right\} \ \rho \ \alpha_
s \left\langle\bar  \psi \psi \right\rangle^ 2 ,} \eeq
\par
\smallskip
\noindent where $ \rho $ is the parameter controlling the deviation from the
vacuum saturation assumption. The sum-rules analysis of the vector and
axial-vector channels gives \cite{SNB}:
\medskip
\beq{ \rho \ \alpha_ s \left\langle\bar  \psi \psi \right\rangle^ 2
\simeq  (3.8 \pm  2.0)\ 10^{-4}{\mbox GeV}^6 , } \eeq
which signals a large deviation from the na\"\i ve
vacuum saturation assumption.
\par
\smallskip
\nin $ {\bf - \ \left\langle O_8 \right\rangle } $ {\bf are the dimension-8
operators } whose strengths are poorly known as they
involve large numbers of operators. Some of their effects have been
calculated in Ref. \cite{BRO} and have been estimated in Ref. \cite{BRA2}
to be about 10$^{-5}$.
\par
\medskip
We are now in a position to write down the theoretical QCD expression of $
 R_\tau $
in the form
\smallskip
\beq{ R_\tau   \equiv  3 \left( \left\vert V_{ud} \right\vert^ 2+
\left\vert V_{us} \right\vert^ 2 \right)\ S_{EW} \left\{ 1 + \delta_{ EW}+
\delta^{ (0)}+ \sum^{ }_{ D=2,4...}\delta^{ (D)} \right\} \ , } \eeq
\par
\smallskip
\noindent where $ \left\vert V_{ud} \right\vert
\simeq  0.9753 \pm  0.0006$, and
$
 \left\vert V_{us} \right\vert  \simeq  0.221 \pm  0.003 $ are the CKM
mixing angles.
\medskip
\noindent
\subsection{The size of the different corrections}
\par
\smallskip
\noindent The different corrections are : \par
\smallskip
\noindent {{\bf $-$ \ Electroweak}} \par
\smallskip
\nin
$ S_{EW}= 1.0194 $ from summing,
via the RGE, the leading ``log contribution" \cite{MARC}.
 \par
\smallskip
\nin
$ \delta_{ EW}= \displaystyle{ 5 \over 12}\ {\alpha \over \pi}  = 0.0010 $ is
the next-to-leading electroweak contribution \cite{LI}.
\par
\smallskip
\noindent  {\bf $-$ \ {Perturbative}}
\par
\smallskip
\nin
We use the available calculations from the $e^+e^- \rar $ hadron
process \cite{KATA}, from which we deduce \cite{BRA2,LEDI}:
\beq{ \delta^{ (0)}  = {\alpha_ s \over \pi}  + 5.2023\
\left({\alpha_ s \over \pi} \right)^2+ 26.366\ \left({\alpha_ s \over \pi}
\right)^3+(78\pm  50) \left({\alpha_ s \over \pi} \right)^4 , } \eeq
\par
\smallskip
\noindent where, for a typical value of $ \alpha_ s \left(M_\tau \right) =0.35
$, they are respectively 11.1, 6.5, 3.6 and $(1.1\pm 0.7)$ \%,
of the leading-order term. The error has been estimated using an
algebraic growth of the coefficient $(50 = 2K_3(K_3/K_2)$, where in the
$\overline{MS}$-scheme $K_2 =1.6398$ and $K_3=6.3711$ are
the coefficients of the $\alpha_s^2$ and $\alpha^3_s$-terms
of the D-function
obtained in
\cite{KATA}. The error has been multiplied by a factor 2 in order to
be more conservative.
The convergence of the
perturbative series for $ \alpha_ s\geq  0.35 $ has been improved in Ref.
\cite{LEDI}
 after a
resummation of the series and by using an expansion other than the
$\alpha_s$ one used in (20). Such a modified expansion appears to be less
dependent on the subtraction point $\mu$ and on the choice of the
renormalization scheme. For a typical value of $\delta^{(0)} \simeq
0.22$, these different uncertainties in the perturbative series induce
an error of about .0017 for the value of $\alpha_s(M^2_Z)$.
 \par
\smallskip
\noindent  {\bf $-$ \ Quark mass}
\par
\smallskip
\nin
Using the quark mass values
quoted in (12), one has
\beq
 \delta^{ (2)}\simeq
-(0.7 \pm  0.2) \%,
\eeq
which comes mainly from $ \hat  m_s. $ \par
\smallskip
\noindent  {\bf $-$ \ Non-perturbative}
 \par
\smallskip
\nin
$ \delta^{ (4)}= -(8\pm 1)10^{-3} $ using (15), where one should notice that
 due to the
$s$-structure of
$ R_\tau $ and the Cauchy theorem, the leading-order contribution of the $
\left\langle O_4 \right\rangle $
effect is zero,
explaining the small value of $ \delta^{ (4)} $, which is only
induced by
the radiative corrections responsible for the $s$
-dependence of $ \left\langle O_4
\right\rangle $. \par
\smallskip
\nin
$ \delta^{ (6)}\simeq  -(7\pm 4)10^{-3} $ using (18). The relative
smallness of this contribution is due not only to the $1/M_\tau^6$
suppression, but also to some compensation between the vector and
axial-vector contributions in $R_\tau$. This nice compensation also
happens to order $\alpha_s$\cite{ADAM},
where the remaining radiative corrections
are much smaller than the errors in (18).
\par
\nin
$ \delta^{ (8)}\simeq  10^{-5}
$ using the vacuum-saturation estimate of
the calculated contributions.
\par
\nin
Adding these different
non-perturbative contributions, one obtains:
\beq
\delta_{SVZ} \equiv \sum_{D=2}^{8} \delta_D  \simeq -(15 \pm 5)10^{-3}
\eeq
One can remark that the sum of the non-perturbative
effects is tiny as it is of the order of the estimated perturbative error.
This is mainly due to the vanishing of the $ \left\langle O_4 \right\rangle $
effects to leading order
and to the fact that operators of dimension $ D\geq 6 $ effects are highly
suppressed in powers of $ 1/M_\tau . $
\subsection{\bf Exotic contributions beyond the SVZ-expansion}
\nin
 From the theoretical point of view, one has also made
some progress (which should be pursued)
in the understanding of some eventual ``exotic effects"
not contained in the SVZ-expansion. Instanton-like effects, though not
under good control,
have been shown to be negligible, as they induce
a correction in the range $10^{-6}$--$10^{-3}$ \cite{NASON}, which is
much smaller
than the non-perturbative effects within the SVZ-expansion retained
previously.
The most dangerous effect
might be due to the eventual existence of dimension-two operators, which
might appear in the massless limit and
which are
not contained
in the original SVZ-expansion as argued by Altarelli \cite{ALT}:

\nin
{\it I think it is a fair statement that there is no theorem that
guarantees the absence of $1/M^2_\tau$ terms in $R_\tau$ in the massless
limit by proving that terms of order $(\Lambda/M_\tau)^2$ cannot arise...
}

\nin
 Phenomenological constraints on this
term (assuming its existence)
from QCD spectral sum rules analysis of
the $e^+e^-$ data \cite{SN2} provide an estimate of about
-$(0.5 \pm 3.1)$\%
correction and indicates that this term (if there) contributes as an
imaginary mass. However, this constraint is not
strong enough for excluding radically
this possibility. Some constraints of this type should be derived
from other sources.
 Morever, the absence of these terms
have been shown in a formal way by \cite{AMU}, \cite{ZAK} using
arguments based on U.V {\it renormalon}. The proof is well
summarized by F. Wilczek \cite{WIL} as:

\nin
{\it Mueller has given an important, although not entirely rigorous,
argument that no $\Lambda^2/Q^2$ term can appear. The argument is
a little technical, so I won't be able to do it full justice here but I
will attempt to convey the main idea.
The argument is based on the idea that at each successive power of 1 over
$Q^2$ one can make the perturbation series in QCD, which is a badly
divergent series in general, at least almost convergent, that is Borel
summable, by removing a finite number of obstructions. Furthermore the
obstructions are captured and parametrized by the low dimension operators
mentioned before. Once these obstructions are removed, the remaining
(processed) perturbation expression converges on the correct result for
the full theory. Neither in the obstructions nor in the residual
perturbative expression do the potentially dangerous terms occur--which
means that they don't occur at all.}

\nin
But, Altarelli \cite{ALT} continues with another statement:

\nin
{\it I also stress that the advantage from the absence of $1/M^2_\tau$
terms in $R_\tau$ could be an illusion when comparing $\alpha_s(M_\tau)$
measured from $R_\tau$ with $\alpha_s(Q)$ derived from some other
process, because one needs control of $\alpha_s(Q)$ down to terms
of order $\Lambda^2/M^2_\tau$, while only the asymptotic form of
$\alpha_s(Q)$ is known. For example, any freezing mechanism at $Q \simeq
\Lambda$ introduces typical corrections of order $\alpha^2_s(M^2_\tau)
\Lambda^2/M^2_\tau$.}

\nin
I have worked out explicitly a check of this statement, by using
the following  expression
$1/\log((Q^2+C^2)/\Lambda^2)$,
of $\alpha_s$ at
low-energy,
instead of the usual asymptotic
 $1/\log(Q^2/\Lambda^2)$-behaviour at high-energy.
 Using the generous range $\Lambda \leq C \leq M_\rho$, where $M_\rho$
 is a typical hadronic scale,
this effect, which
is an $\alpha^2_s$
effect, induces a correction less than
$5 \times 10^{-3}$ in the tau-decay rate, which is about the same as the
error from the non-perturbative effects in (21).

\nin
Then, one can conclude, {\it without any doubts},
 that the different non-perturbative effects
within or beyond the SVZ-expansion are tiny and make the $\tau$-decay
a prime place for determining $\alpha_s$.
\begin{table}
\begin{center}
\begin{tabular}{|c|c c c  c|}
\hline
$\alpha_s(M^2_{\tau})$&$R_{\tau,V}$
& $R_{\tau,A}$ & $R_{\tau,S}$ & $R_\tau$ \\
\hline
$0.16$ & $1.59 \pm 0.02$ & $1.49 \pm 0.03$ & $0.145 \pm 0.004$ &
$ 3.23 \pm 0.01 $\\
$0.18$ & $1.61 \pm 0.02$ & $1.51 \pm 0.03$ & $0.145 \pm 0.004$ &
$ 3.26 \pm 0.01 $\\
$0.20$ & $1.62 \pm 0.02$ & $1.53 \pm 0.03$ & $0.145 \pm 0.005$ &
$ 3.29 \pm 0.01 $\\
$0.22$ & $1.64 \pm 0.02$ & $1.54 \pm 0.03$ & $0.145 \pm 0.005$ &
$ 3.33 \pm 0.02 $\\
$0.24$ & $1.66 \pm 0.02$ & $1.56 \pm 0.03$ & $0.145 \pm 0.005$ &
$ 3.37 \pm 0.02 $\\
$0.26$ & $1.68 \pm 0.02$ & $1.58 \pm 0.03$ & $0.145 \pm 0.005$ &
$ 3.41 \pm 0.02 $\\
$0.28$ & $1.70 \pm 0.02$ & $1.61 \pm 0.03$ & $0.145 \pm 0.005$ &
$ 3.45 \pm 0.02 $\\
$0.30$ & $1.72 \pm 0.02$ & $1.63 \pm 0.03$ & $0.145 \pm 0.006$ &
$ 3.50 \pm 0.02 $\\
$0.32$ & $1.75 \pm 0.02$ & $1.65 \pm 0.03$ & $0.145 \pm 0.006$ &
$ 3.54 \pm 0.03 $\\
$0.34$ & $1.77 \pm 0.02$ & $1.67 \pm 0.03$ & $0.145 \pm 0.006$ &
$ 3.58 \pm 0.03 $\\
$0.36$ & $1.79 \pm 0.02$ & $1.69 \pm 0.03$ & $0.144 \pm 0.006$ &
$ 3.63 \pm 0.03 $\\
$0.38$ & $1.81 \pm 0.03$ & $1.71 \pm 0.03$ & $0.144 \pm 0.007$ &
$ 3.67 \pm 0.04 $\\
$0.40$ & $1.83 \pm 0.03$ & $1.73 \pm 0.03$ & $0.143 \pm 0.007$ &
$ 3.71 \pm 0.04 $\\
$0.42$ & $1.85 \pm 0.03$ & $1.75 \pm 0.04$ & $0.143 \pm 0.007$ &
$ 3.75 \pm 0.04 $\\
$0.44$ & $1.87 \pm 0.03$ & $1.77 \pm 0.04$ & $0.142 \pm 0.008$ &
$ 3.79 \pm 0.04 $\\
\hline
\end{tabular}
\end{center}
\caption{ QCD predictions [2,3,5] for the different components
of the $\tau$ hadronic width.}
\end{table}
\par
\smallskip

\subsection{\bf{The value of} \boldmath{$\alpha_s$}}
By confronting the QCD predictions in Table 1 with the data in (5),
one deduces
\medskip
\beq{ \alpha_ s \left(M^2_\tau \right)  = 0.36 \pm  0.03.} \eeq
\par
\smallskip
\noindent We run this value at $M_\tau$
up to $ M_Z, $ by using the matching conditions
at the heavy quark
thresholds \`a la \cite{BERN}, i.e at the value of the running
c and b quark masses, which we
deduce from the QCD spectral sum rule estimates of the
$perturbative$ pole masses \cite{MASS}:
\beq
M_c(p^2 =M^2_c) \simeq 1.45 \pm 0.05~\mbox{GeV}~~~~~~~~~~~
M_b(p^2 =M^2_b) \simeq 4.58 \pm 0.05~\mbox{GeV}.
\eeq
One should notice that the values of the pole masses given above
come from the standard relativistic sum rule estimate through the
perturbative
euclidian mass, such that they are not affected by renormalon-type
contributions which induce a non-perturbative effect of the order
$\Lambda$ in
the pole mass used in non-relativistic
sum rule and heavy quark effective theory. Such an effect makes these
non-relativistic masses slightly larger as in the potential models
but their definiton is still ambiguous at present.

\nin
The previous matching procedure
already includes in it \cite{BERN} the tiny $(\alpha_s/\pi)^2$
effects of about
$\delta^{(H)} \simeq 5~10^{-4}$ from
virtual
heavy quark loops obtained in \cite{CHET}.
At the end of the day, one obtains:
\medskip
\beq{ \alpha_ s \left(M^2_Z \right)  \simeq  0.122 \pm  0.003 \pm 0.001
\ \ ,
} \eeq
where the matching procedure has induced the last error which is
a conservative error.
This result is
in nice agreement with,
and slightly more precise than, the present LEP
average \cite{BET}
 $ \alpha_ s \left(M^2_Z \right) \simeq   0.125 \pm  0.005 $,
done without including the $ \tau $-decay source. This precision indicates that
a
modest accuracy at $ M_\tau $ leads to a high-precision measurement at $ M_Z $
as the
error bars run like $ \alpha^ 2_s $. The agreement between the
independent determinations
at $ M_\tau $
and $ M_Z $ also shows that $ \alpha_ s $ runs as expected in a QCD
asymptotically-free theory.
 \par
\medskip
\subsection{\bf {ALEPH test from the weighted-moment distributions}}
\medskip
The ALEPH collaboration at LEP
\cite{ALEPH} has tested the previous result
by working with
the weighted moments distributions \cite{LEDI}:
\medskip
\bea
R^{k\ell}_ \tau   &\equiv& \int^{ M^2_\tau}_0\ ds\ \left(1-{s \over
s_0} \right)^k \left({s \over M^2_\tau} \right)^{\ell} \
 {dR_\tau \over ds}  \nnb \\
 D^{k\ell}_ \tau
 &\equiv &R^{k\ell}_ \tau \over R^{00}_\tau,
\eea
\noindent
which are sensitive to $\alpha_s$ and to the non-perturbative
condensates. These moments
have the advantage of being directly measurable, thanks to the hadronic
invariant-mass squared distribution $ {dR_\tau / ds} $. The factor $
\left(1-{s / M^2_\tau
} \right)^k $ supplements $ \left(1-{s / M^2_\tau}
\right)^2 $
and squeezes the integrand near the positive real
 axis.
This improves the reliability of the OPE and of the analysis.
Using the present LEP data,
which are still
statistically limited,  one obtains from a 4-parameter fit of 5
observables ($R_\tau, \ D^{1l}_\tau \ l$=0--3):
\medskip
\bea \alpha_ s \left(M_\tau \right) & =& (0.34\pm 0.04) \nnb \\
\left\langle \alpha_ sG^2 \right\rangle &  =& (0.06\pm 0.05)\ {\mbox
GeV}^4 \nnb \\
\rho \ \alpha_ s \left\langle\bar  \psi \psi \right\rangle^ 2 & =& (4\pm
4)10^{-4}~\mbox{ GeV}^6  \nnb \\
\langle O_8 \rangle &=& (3 \pm 2) 10^{-3}~ {\mbox GeV}^8,
\eea
\par
\smallskip
\noindent where the results are strongly correlated.
The value of $ \alpha_s
$ is compatible
with the previous ones from the decay modes. The values of the condensates
are still inaccurate, but they
 are compatible with the ones from QSSR analysis
of charmonium and $ e^+e^- \longrightarrow  I=1 $ hadron data used
previously. However, due to these strong correlations, it will be
also useful and it is possible
to extract with a much better accuracy
the ratios of different condensates in order to also test the sum rule
predictions of these quantities.
Adding these different correlated
non-perturbative contributions, one also obtains:
\beq
\delta_{SVZ} \simeq  (3 \pm 5)10^{-3},
\eeq
which is consistent with (22) and confirms
the smallness of the non-perturbative contributions in $R_\tau$.
The
experiments which can produce millions of $\tau$ such as the
 $\tau$C/B-factory
are the
best place for
improving these interesting results on $\alpha_s$ and on the size
of the QCD condensates.

\section{ \boldmath{${ \alpha}_ {s} $}{\bf\ FROM THE SUM OF EXCLUSIVE
MODES}}\par
\nin
This section is based on the work in \cite{PICH}.
\par
\subsection{\bf {The vector channel} }\par
\smallskip
It is known that using an $ SU(2)_I $
 rotation, one can relate the
vector component of the $ \tau $-decay with the $ e^+e^- \longrightarrow  I=1
$ hadrons data :
\medskip
\bea
 R_{\tau ,V}&\equiv&
   {\Gamma \left(\tau  \longrightarrow  \nu_ \tau V \right)
\over \Gamma \left(\tau  \longrightarrow  \nu_ \tau e\bar  \nu e \right)}
\nnb \\
  &  = &{3\cos^2 \theta_c \over 2\ \pi \alpha^ 2}\ S_{EW}\ \int^{
M^2_\tau}_ 0\ {ds \over M^2_\tau} \ \left(1 - {s \over M^2_\tau} \right)^2
\left(1 + {2s \over M^2_\tau} \right)\ s\ \sigma^{ I=1}_{e^+e^-
\longrightarrow  V_0}(s)\ .
\eea
\par
\smallskip
\medskip

\noindent The values of $ R_{\tau ,V} $ from the $ \tau $-data and estimated
from (19) are given in
Table 2,
\noindent where one should notice that the estimates of $ K^-K^0 $ and $ \pi^
-K^+K^- $ have been
done using $SU(3)$
rotations with the appropriate phase space factor. By combining
the data and the estimated results, one can obtain the best value :
\medskip
\beq{ R^{exp}_{\tau ,V}  = 1.78 \pm  0.03\ \ . } \eeq
\par
\smallskip
\noindent Comparing this result with the QCD expression :
\medskip
\beq{ R_{\tau ,V}  = {3 \over 2} \left\vert V_{ud} \right\vert^ 2
 \left(1+\delta^{
(0)}+ \sum^{ }_{ D=2,4,...}\delta^{ (D)}_{ud,V} \right), } \eeq
\par
\noindent where
\bea
\delta^{ (2)}_{ud,V}&\simeq&  -(0.6\pm 0.2)10^{-3} ,\nnb \\
\delta^{(4)}_{ud,V}&\simeq&  (0.8\pm 0.3)10^{-3} , \nnb  \\
\delta^{ (6)}_{ud,V} & \simeq&  (2.4\pm 1.3)10^{-2} ,
 \eea
\begin{table}
\begin{center}
\begin{tabular}{|c|c|c|c|}
\hline
$V^-$&$\tau$-data & $e^+e^-$ & $e^+e^-$ \\
  &  & (Ref. \cite{EID}) & (our estimate) \\
\hline
$\pi^-\pi^0$ & $1.355 \pm 0.021$ & $1.349 \pm 0.046$ & $1.346 \pm 0.040$
\\
$2\pi^-\pi^+\pi^0$ & $0.307 \pm 0.013$ & $0.248 \pm 0.015$ &
$0.268 \pm 0.040$ \\
$\pi^-3\pi^0$ & $0.063 \pm 0.009$ & $ 0.061 \pm 0.003$ & $0.057 \pm 0.010
$ \\
$\pi^-\omega$ & $0.090 \pm 0.028 $ & $0.128 \pm 0.018$ & $0.129\pm 0.023$
\\
$3\pi^-2\pi^+\pi^0$ &$ 0.003 \pm 0.001 $& --&-- \\
$(6\pi)^-$ & - & $0.011 \pm 0.002$ & $0.008 \pm 0.003$ \\
$\pi^-\pi^0\eta$ & $ 0.010\pm 0.002$
 & $0.007 \pm 0.001$ & $0.008 \pm 0.003$ \\
$K^-K^0$ &$ \leq 0.015$ & $0.006 \pm 0.002$ & $0.009 \pm 0.001$ \\
$\pi^-K^-K^0$ & $0.011 \pm 0.005$ & -- & $0.009 \pm 0.003$ \\
\hline
\hline
$R_{\tau,V}$ & $1.768 \pm 0.032$ & $1.693 \pm 0.049$ &
$1.725 \pm 0.069$\\
\hline
\end{tabular}
\end{center}
\caption{ Contributions of different exclusive $\tau$-decay
modes $\tau^- \rar \nu_\tau V^-$ to $R_{\tau,V}$.}
\end{table}
\par
\smallskip
\noindent one should notice that the strengh of the $ \left\langle O_6
\right\rangle $ effect is larger here than in the inclusive mode
$ R_\tau $. In addition to the inaccuracy of the exclusive data,
this fact limits the accuracy on the determination of $ \alpha_ s $ from
 the vector channel. We deduce from Table 2:
\medskip
\beq \alpha_ s \left(M^2_\tau \right)  \simeq  0.35\pm 0.05 .
\eeq
\par
\medskip
\noindent {\bf 3.2. The axial-vector channel} \par
\smallskip
We give in Table 3 the exclusive decays in the
axial-vector channel and their sum, using the more recent data
quoted in \cite{FERN}.
The QCD expression of $ R_{\tau ,A} $ is
\smallskip
\beq{ R_{\tau ,A}  = {3 \over 2} \left\vert V_{ud} \right\vert^ 2
\left(1+\delta^{ (0)}+ \sum^{ }_{ D=2,4,...}\delta^{ (D)}_{ud,A} \right) .
 } \eeq
\begin{table}
\begin{center}
\begin{tabular}{|c|c|}
\hline
$A^-$ & $R_{\tau \rar A^-}$  \\
\hline
$\pi^-$ & $0.660 \pm 0.020 $ \\
$2\pi^-\pi^+$ & $0.467 \pm 0.017$ \\
$\pi^-2\pi^0$ & $0.499 \pm 0.019$ \\
$3\pi^-2\pi^+$ & $0.005 \pm 0.001$ \\
$2\pi^-\pi^+2\pi^0$ & $0.027 \pm 0.003$ \\
$\pi^-4\pi^0$ & $0.008 \pm 0.004$ \\
\hline
\hline
$R_{\tau \rar A^-}$ & $ 1.666 \pm 0.033$ \\
\hline
\end{tabular}
\end{center}
\caption{ Contributions of different exclusive $\tau$-decay
modes $\tau^- \rar \nu_\tau A^-$ to $R_{\tau,A}$.}
\end{table}
\par
\smallskip
\noindent The non-perturbative corrections are :
\medskip
\bea  \delta^{ (2)}_A & \simeq &-(1\pm 0.2)10^{-3} \nnb \\
\delta^{
(4)}_A & \simeq &-(4.6\pm 0.7)10^{-3} \nnb  \\
 \delta^{ (6)}_A & \simeq&
-(3.8\pm 2.0)10^{-2} . \eea
\par
\smallskip
\noindent A comparison of the data and of $ R_{\tau ,A}^ {QCD}$ in Table
1 leads to :
\medskip
\beq \alpha_ s \left(M^2_\tau \right)  \simeq   0.34\pm 0.05,
\eeq
\par
\smallskip
\noindent where the central value is slightly
lower than the one from the inclusive and
vector channels,
although consistent. This slightly
lower value of $\alpha_s$ still signals
a remaining though small deficit in this exclusive channel, but compared
with the previous data used in \cite{PICH}, the new data \cite{FERN} have
provided an improvment of the value of $\alpha_s$ from
this axial channel.
A more precise value of $\alpha_ s $ from
 this
channel needs a better measurement of the $3\pi$ and of some other
multipion channels.

\begin{table}
\begin{center}
\begin{tabular}{|c|c|}
\hline
$S^-$ & $R_{\tau \rar S^-}$  \\
\hline
$K^-$ & $0.043 \pm 0.004$ \\
$K^{*-}(\geq 0\pi^0)$ & $0.081 \pm 0.010$ \\
$K^{*0}\pi^-(\geq 0\pi^0)$ & $0.021 \pm 0.010$ \\
\hline
\hline
$R_{\tau,S}$ & $0.145 \pm 0.015$ \\
\hline
\end{tabular}
\end{center}
\caption{ Contributions of different exclusive $\tau$-decay
modes $\tau^- \rar \nu_\tau S^-$ to $R_{\tau,S}$.}
\end{table}
\nobreak\ \nobreak\  \par
\medskip
\noindent {\bf 3.3. The Cabibbo-suppressed channel} \par
\smallskip
\nin
We show the data in Table 4, where we have used the most recent
data for $K^-$ \cite{FERN}, with an improved error by about a factor 3.
\par
\nin
The QCD expression is :
\beq R_{\tau ,S}  = 3 \left\vert V_{us} \right\vert^ 2
\left(1+\delta^{ (0)}+ \sum^{ }_{ D=2,4}\delta^{ (D)}_{us} \right),
\eeq
\par
\smallskip
\noindent which predicts from Table 1:
\medskip
\beq{ R^{QCD
}_{\tau ,S}  \simeq  0.145\pm 0.006 .} \eeq
\par
\nin
One should notice that the estimate is ``almost" insensitive to the
value of $ \alpha_ s $. This is mainly due to the ``almost"
cancellation of the
$ \alpha_ s $ and
$ m^2_s $ contributions, while the higher dimension condensates
``almost" cancel
with the $ \alpha^ 2_s $ and $ \alpha^ 3_s $ effects. These different
 cancellations
make $ R^{QCD}_{\tau ,S} $ to be ``almost"
equal to the prediction of the na\"\i ve quark-parton model :
\medskip
\beq{ R^{naive}
_{\tau ,S}  = N_c \left\vert V_{us} \right\vert^
2\simeq  0.147 . } \eeq
\par
\smallskip
\noindent At present, one cannot extract useful informations on the
structure of QCD from the data. But a
high-accuracy measurement of this channel can provide a
measurement of
the strange quark mass or a constraint on some exotic
$ D=2 $ ``operator" not contained in the SVZ-expansion.
 \par
\medskip
\subsection{The sum of the exclusive modes} \par
\smallskip
\noindent$ R^{exclusive}_\tau $ can be obtained by adding (30) to Tables 3
 and 4.
In this way one
obtains :
\beq{ R^{exclusive}_\tau   = 3.59\pm 0.05, } \eeq
which leads to :
\beq{ \alpha_ s \left(M^2_\tau \right)  \simeq  0.34\pm 0.04 , } \eeq
in good agreement with the one
from the inclusive mode in (23), though the central value in (41) is
slightly
lower.
\section{ STABILITY TEST FROM
  e$^ {+}$ {e}$^ {-}$
DATA}
\medskip
We \cite{PICH}
test the stability of the previous result by varying the $ \tau $-mass. In
so doing, we use (29) for arbitrary values of $ M_\tau \equiv  M $ and we
use the $ e^+e^- \longrightarrow  I=1 $
hadron data. Our result is shown in Fig. 2.
The bars come from the $ e^+e^- $ data. The continuous line is the fit
 for $
\alpha_ s \left(M_\tau \right) = 0.33 $, $ \delta^{ (6)}_V= 0.024 $
and $ \delta^{ (8)}_V= -0.010 $.
The shaded region shows the effect of the errors in $ \delta^{ (6)}_V $,
which are $ \pm 0.013. $
The hatched regions show the effects of the errors in $ \alpha_ s
\left(M^2_\tau \right) $, taken to
be $ \pm 0.03 $ at fixed values of the condensates. It is clear that there is
a
good agreement between the theory and the data,
 except at low $ M $ where the
role of the higher dimension $ D\ge 8 $ condensates is important as it changes
completely the predicted behaviour below $ 1.2 $ GeV.  One can impose an
 agreement
of the theory with the data until 1 GeV by fitting the value of the $D=8$
 operators.
One should however
notice that the ratio of the
obtained value of $ \delta^{ (8)}_V $
 over the $ D=6 $
corrections is 1.25,
 signaling presumably the breaking of the OPE at such a
low scale of   1 GeV.
Here, I should also mention that Ref.
\cite{TRUONG} has also used the $e^+e^-$data in order to
test the accuracy of the estimate of $\alpha_s$ from $\tau$-decay.
However, after a careful reading and check of the method used there,
one can realize that the analysis emphasizes the region above 1.8 GeV
where the data are in contradiction when available. Indeed, the author
works with a difference of two finite energy sum
rules of radius $M^2_\tau$ and $s_0$. Moreover, one can also notice
that in the uses of the
usual FESR
and dispersion relation for the D-function, the results depend
strongly on the way the parametrization of the data in the energy
region above 1.8 GeV
is done.
That is due to the well-known sensitivity of these methods
on the medium-energy behaviour of the spectral
function because of the usual $s^n$ weight factors entering in
the sum rules.
 Fortunately, this is not the case of $R_\tau$ thanks to the
$(1-s/M^2_\tau)^2$ threshold effect weight factor
which suppresses this source
of uncertainty.
One can fairly conclude
that the analysis done in \cite{TRUONG} has nothing
to do with the estimate of $\alpha_s$, but instead, only shows the
already known
 evidence of the unstability of the results from
 the FESR and usual dispersion
relation approaches due to the medium-energy behaviour of the
spectral function. One could instead consider this analysis
 as a test of
the validity of different parametrizations used in this
medium-energy regime and how fast they reach the asymptotic regime
of QCD. However,
a more definite conclusion needs a careful
inclusion of the error bars from the different data parametrizations.
\medskip
\begin{table}
\begin{center}
\begin{tabular}{|c|c|}
\hline
Observables &  $\alpha_s(M^2_\tau)$ \\
\hline
$R_\tau$ & $0.36 \pm 0.03$ \\
$D^{kl}_\tau$ &$0.330 \pm 0.046$ \\
$R_{\tau,V}$& $ 0.35 \pm 0.05$ \\
$R_{\tau,A}$& $ 0.34 \pm 0.05$ \\
$R_{\tau}^{exclusive}$& $ 0.34 \pm 0.04$ \\
\hline
\hline
$average$&$0.347 \pm 0.030$ \\
\hline
\end{tabular}
\end{center}
\caption{ Values of $\alpha_s$ from different observables in
 $\tau$-decays.}
\end{table}
\section{\bf CONCLUSION}
We have reviewed the different determinations of the QCD coupling
$\alpha_s$ from the inclusive \cite{BRA}-\cite{PICH}
and exclusive \cite{PICH}
$ \tau $-decay data, which we summarize in Table 5, from which one
can deduce:
\beq
\alpha_s(M^2_\tau) \simeq 0.347 \pm 0.030
\ \Longrightarrow \
\alpha_s(M^2_Z) \simeq 0.121 \pm 0.003 \pm 0.001,
\eeq
where the last
error quoted here at $M_Z$ is a conservative error induced by the
one of the heavy quark masses and by
the
procedure of the
matching conditions at the quark thresholds \cite{FRAN}.
We compare this
result with the ones from the other determinations
(see Figs. 3 and 4 which
are updated from \cite{WIL, BET} as we have included the new average
of $R_\tau$ and the global fit from the elecroweak data from LEP).
As one can see in these figures,
it turns
 out that $
\tau
$-decays are a good and presumably one of
the best place for accurately measuring
$\alpha_s$. The agreement with the other LEP results at $M_Z$ is a
strong indication of the $1/\log$-running of $\alpha_s$ as expected
from the asymptotically free theory of QCD. The ALEPH \cite{ALEPH}
measurement of the condensates also constitutes
a test, in a method-independent way, of the underlying
non-perturbative aspects of QCD within and beyond
the SVZ expansion. Indeed, one can also use this result the other way
around: the accurate agreement between the values of $\alpha_s$ obtained
from $\tau$-decay and from other high-energy processes such as LEP, is
evidence against the possible
existence of huge exotic non-perturbative effects beyond
the SVZ-expansion. At present,
 the errors
 in
the determinations of these different QCD
{\it fundamental} or more properly {\it
universal} parameters are dominated by the
 statistical
errors,
 which can be notably reduced in the future  $\tau$C/B-factory
 experiments.

\nin
As I started my talk with some quotations,
I shall conclude it in the same way:

\nin
{\it I went into some detail into the analysis of tau decay
because I think it's not only important in itself but quite fundamental,
and it connects with many other issues. In particular, this kind of
argument could potentially provide a rigorous foundation for the QCD or
ITEP sum rules which are the basis of a very successful phenomenology}...
(F. Wilczek \cite{WIL}).

\section*{Acknowledgements}
This talk is based on the works done in collaboration with Eric Braaten
and Antonio Pich.
 I want to express my
gratitude to the organizers of the Fubini symposium and of the
QCD--LEP meeting for
these quite convivial (and likely familial) meetings.
\section*{Figure captions}
Fig. 1 : Integration contour in the complex s-plane, used to obtain (10).

\nin
Fig. 2 : $R_{\tau,V}$ in (29) as function of $M \equiv M_\tau$.

\nin
Fig. 3 : Different sources for evaluating $\alpha_s$ at different
energies.

\nin
Fig. 4 : Different values of $\alpha_s$ from Fig. 3, but evaluated at
$M_Z$.
\par
\vfill\eject


\medskip
\noindent
\begin{thebibliography}{999}
\bibitem{BRA}
E. Braaten, {\it Phys. Rev. Lett.}
{\bf 60} (1988) 1606, {\it Phys. Rev.} {\bf D39} (1989)
1458, {\it Proc. $\tau$Cf Workshop, Stanford} (1989), ed. L.V. Beers;
S. Narison and A. Pich,{\it Phys. Lett} {\bf B211} (1988) 183;
A. Pich, {\it Proc. $\tau$Cf
workshop, Stanford} (1989), ed. L.V. Beers ; {\it Proc. Workshop on
$\tau$-physics, Orsay}
(1990), eds. M. Davier and B. Jean-Marie (Ed. Fronti\`eres);
{\it Proc. $\tau Cf$ workshop, Marbella} (1993);
F. Le Diberder, {\it Proc. $\tau Cf$ workshop, Marbella} (1993);
 S. Narison, {\it Proc. $\tau Cf$ workshop, Marbella} (1993).
\bibitem{BRA2}
E. Braaten, S. Narison and A.
Pich, {\it Nucl. Phys.} {\bf B373} (1992) 581.
\bibitem{LEDI}
F. Le Diberder and A. Pich,
{\it Phys. Lett.} {\bf B286} (1992) 147 and {\bf B289}
(1992) 165.
\bibitem{ALEPH}ALEPH collaboration: D. Buskulic et al., {\it Phys. Lett.}
{\bf B307} (1993) 209.
\bibitem{PICH}
S. Narison and A. Pich,
{\it Phys. Lett.} {\bf B304} (1993) 359.
\bibitem{BRAT}E. Braaten, {\it Phys. Rev. Lett.} {\bf 71}
 (1993) 1316.
\bibitem{AMU}A. Mueller, {\it QCD-20 years later workshop, Aachen}
(1992).
\bibitem{WIL}F. Wilczek, {\it Lepton-Photon conference, Cornell, Ithaca,
N.Y.} (1993).
\bibitem{SVZ}
M.A. Shifman, A.I. Vainshtein and V.I.
Zakharov, {\it Nucl. Phys.} {\bf B147}
(1979) 385, 448.
\bibitem{SN1}
 J. Gasser and H. Leutwyler, {\it Phys. Rep.} {\bf 87} (1982) 77;
C.A. Dominguez and E. de Rafael, {\it Ann. Phys.}
{\bf 174} (1987) 372;
S. Narison, {\it Phys. Lett.} {\bf B216} (1989) 191.
\bibitem{SNB}For a recent review, see e.g.:
S. Narison, {\it QCD spectral sum rules,
Lecture notes in physics}, {\bf Vol. 26}
(1989) published by World Scientific.
\bibitem{BRO}
 D.J. Broadhurst and S.C. Generalis, {\it Phys.
Lett.} {\bf B165} (1985) 175.
\bibitem{MARC}
W. Marciano and A. Sirlin, {\it Phys. Rev.
Lett.} {\bf 61} (1988) 1815 and {\bf 56 }
(1986) 22.
\bibitem{LI}
E. Braaten and C.S. Li, {\it Phys. Rev.} {\bf D42}
(1990) 3888.
\bibitem{KATA}S.G. Gorishny, A.L. Kataev and S.A. Larin,
{\it Phys. Lett.}
 {\bf B259} (1991) 144;
see also L.R. Surguladze and M.A. Samuel,
{\it Phys. Rev. Lett.}
 {\bf 66} (1991) 560; {\bf 66} (1991) 2416 (erratum).
\bibitem{ADAM}L.E. Adam and K.G. Chetyrkin, Karlsruhe preprint TTP93-26
(1993).
\bibitem{NASON}P. Nason and M. Porrati, CERN preprint, TH.6787/93 (1993);
I.I. Balitsky, M. Beneke and V.M. Braun, Munich preprint,
MPI-Ph/93-62 (1993).
\bibitem{ALT}G. Altarelli,
{\it QCD-20 Years Later workshop, Aachen} (1992).
\bibitem{SN2}S. Narison,
{\it Phys. Lett.} {\bf B300} (1993)
 293.
\bibitem{ZAK} M. Beneke
and V.I. Zakharov, {\it Phys. Rev. Lett.} {\bf 69} (1992) 2472  and
private communication.
\bibitem{BERN} W. Bernreuther and W. Wetzel, {\it Nucl. Phys.} {\bf
B197} (1982) 228 ; W. Bernreuther, {\it Annals. Phys.} {\bf 151}
(1983) 127 and private communication;
see also: G. Rodrigo and A. Santamaria,
{\it Phys. Lett.} {\bf B313} (1993)
 441.
 A. Peterman, CERN-TH.6487/92 (1992).
\bibitem{MASS}S. Narison,
{\it Phys. Lett.} {\bf B197} (1987)
 405; {\bf B216} (1989) 191.
\bibitem{CHET}K.G. Chetyrkin
{\it Phys. Lett.} {\bf B307} (1993) 169.
\bibitem{BET}
 S. Bethke, {\it Proc. XXVI Int. Conf.
on High Energy Physics, Dallas} (1992) .
\bibitem{FERN} E. Fernandez, {\it EPS conference, Marseille} (1993).
\bibitem{EID}S.I. Eidelman  and V.N. Ivanchenko
, {\it Phys. Lett.} {\bf B257} (1991) 437.
\bibitem{TRUONG}T.N. Truong, Ecole polytechnique preprint,
EP-CPth.A266.1093 (1993); {\it Phys. Rev.} {\bf D47} (1993) 3999.
\bibitem{FRAN}Clarifying discussions with F. Le Diberder, A. Pich
and A. Santamaria.
\end{thebibliography}
\end{document}